\documentclass[aip,amsmath,amssymb,reprint]{revtex4-1}
\usepackage{graphicx}

\usepackage{color}

\begin{document}


\title{Cavity-enhanced Ramsey spectroscopy at a Rydberg-atom--superconducting-circuit interface} 



\author{D. M. Walker}
\author{A. A. Morgan}
\author{S. D. Hogan}\altaffiliation{Author to whom correspondence should be addressed: s.hogan@ucl.ac.uk}
\affiliation{
Department of Physics and Astronomy\\
University College London\\
Gower Street, London WC1E 6BT, United Kingdom
}%


\date{\today}

\begin{abstract}
The coherent interaction of Rydberg helium atoms with microwave fields in a $\lambda/4$ superconducting coplanar waveguide resonator has been exploited to probe the spectral characteristics of an individual resonator mode. This was achieved by preparing the atoms in the 1s55s\,$^3$S$_1$ Rydberg level by resonance enhanced two-color two-photon excitation from the metastable 1s2s\,$^3$S$_1$ level. The atoms then travelled over the resonator in which the third harmonic microwave field, at a frequency of $\omega_{\mathrm{res}}=2\pi\times19.556$~GHz, drove the two-photon 1s55s\,$^3$S$_1\rightarrow$1s56s\,$^3$S$_1$ transition. By injecting a sequence of Ramsey pulses into the resonator, and monitoring the coherent evolution of the Rydberg state population by state-selective pulsed electric field ionization as the frequency of the microwave field was tuned, spectra were recorded that allowed the resonator resonance frequency and quality factor to be determined with the atoms acting as microscopic quantum sensors. 
\end{abstract}


\maketitle 


Hybrid quantum systems comprising gas-phase Rydberg atoms coupled to solid-state superconducting microwave circuits are of interest for a range of applications in quantum information processing~\cite{xiang13a,kurizki15a}. The development of these systems was inspired by proposals for hybrid quantum processors with cold polar ground-state molecules coupled to coplanar waveguide (CPW) resonators integrated into superconducting circuits~\cite{rabl06a}. The use of atoms in Rydberg states with high principal quantum number $n$, offers the opportunity to exploit the large electric dipole moments ($\gtrsim1000\,e\,a_0$ for values of $n\gtrsim50$) associated with microwave transitions between Rydberg states~\cite{raimond01a} to achieve higher single-particle coupling rates than expected for polar molecules. This has motivated theoretical work on the implementation of quantum gates~\cite{pritchard14a,sarkany15a,liao19a}, quantum-state transfer~\cite{patton13a,sarkany18a}, microwave-to-optical photon conversion~\cite{kiffner16a,gard17a,han18a,petrosyan19a}, and studies of new regimes of light-matter interactions~\cite{calajo17a} with Rydberg atoms coupled to superconducting circuits.

The challenges in realizing a hybrid Rydberg-atom--superconducting-circuit interface arise primarily from the susceptibility of Rydberg states to electric fields emanating from the cryogenically cooled superconducting chip surfaces. Studies have been reported in which transitions between Rydberg states were driven by microwave fields propagating in normal metal~\cite{hogan12a} and superconducting~\cite{hermann_avigliano14a,thiele15a} CPWs cooled to low temperatures. These permitted the characterization of stray electric fields and the microwave field distributions above the CPW structures. The results of this work motivated the use of (1) helium (He) atoms in these types of experiments to minimize detrimental effects~\cite{hattermann12a} associated with surface adsorption, (2) microwave transitions between Rydberg states for which the sensitivity to stray electric fields is minimal, or can be minimized through the application of electric, magnetic or microwave dressing fields~\cite{jones13a,booth18a,morgan18a,peper19a}, and (3) CPW resonators with geometries in which the centre conductor can be electrically contacted to the ground planes to minimize charge build up, e.g., $\lambda/4$ CPW resonators. 

Following these guidelines, the first demonstration of an interface between Rydberg atoms and microwave fields in a superconducting CPW resonator was recently reported~\cite{morgan20a}. In this work Rydberg He atoms were coupled to the $\omega_{\mathrm{res}}=2\pi\times19.556$~GHz third-harmonic field in a $\lambda/4$ resonator. This field was resonant with the two-photon $1\mathrm{s}55\mathrm{s}\,^3\mathrm{S}_1\rightarrow 1\mathrm{s}56\mathrm{s}\,^3\mathrm{S}_1$ ($|55\mathrm{s}\rangle\rightarrow|56\mathrm{s}\rangle$) transition, which, because of the similarity in the static electric dipole polarizabilities of the $|55\mathrm{s}\rangle$ and $|56\mathrm{s}\rangle$ states, exhibited a low sensitivity to residual uncancelled stray electric fields of $\lesssim50$~mV/cm within $100~\mu$m of the surface of the superconducting chip. Here we build on this work, and report atom--resonator-field interactions with coherence times up to $0.84~\mu$s, and cavity-enhanced Ramsey spectroscopy of the Rydberg-atom--superconducting-circuit interface. 

In the following an overview of the experimental apparatus is first provided. The results of time-domain measurements of Rabi oscillations in the Rydberg state population, arising from the coherent interaction with the resonator field, are then presented. This is followed by a discussion of the cavity-enhanced Ramsey spectroscopy performed in the frequency domain. The measured Ramsey spectra have been compared to the results of numerical calculations of the time-evolution of the Rydberg state population in the presence of the resonator field, to allow the spectral characteristics of the resonator mode to be determined from the experimental data. 


\begin{figure}
\includegraphics[width=0.38\textwidth]{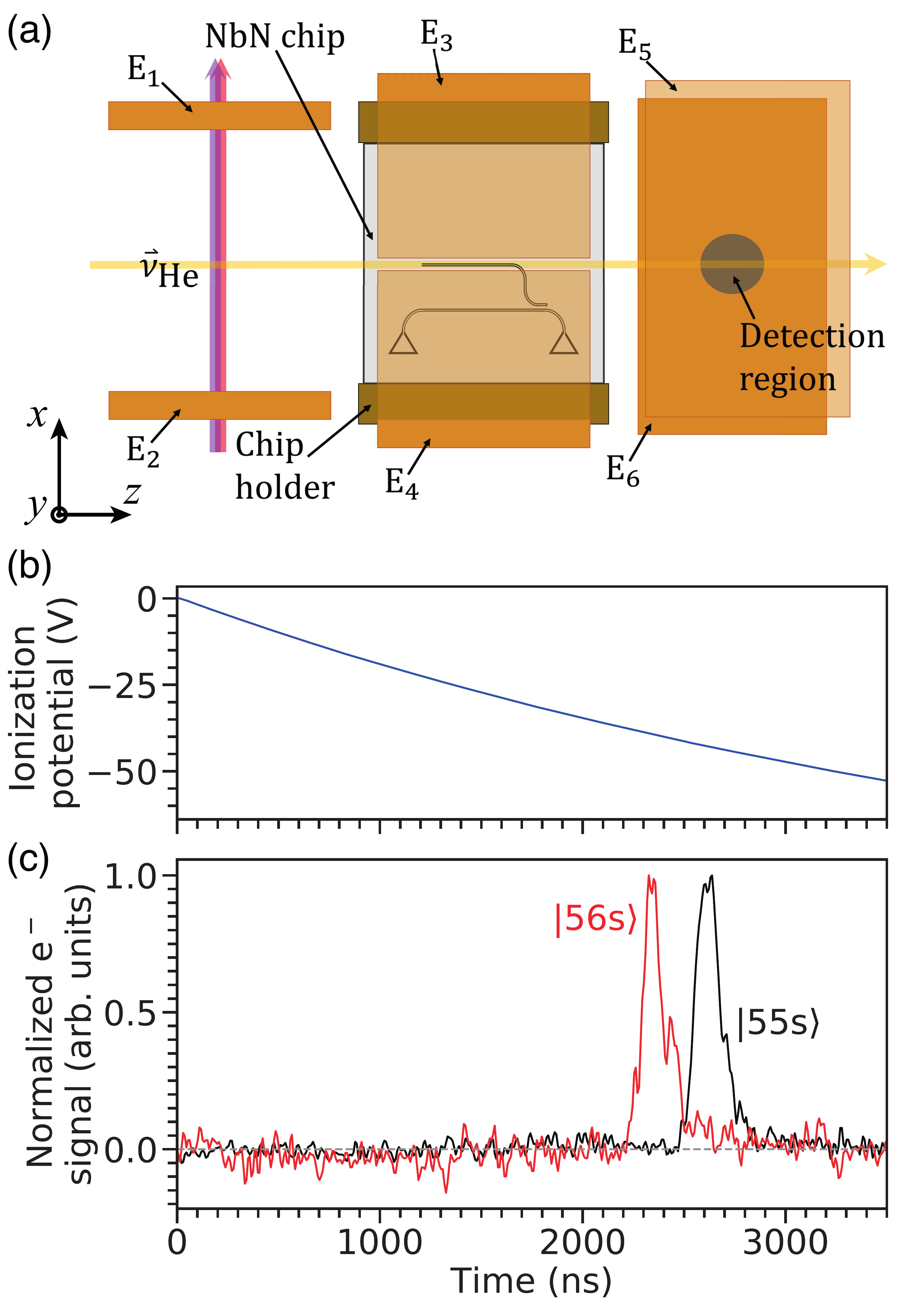}
\caption{(a) Schematic diagram of the cryogenically cooled interaction region in the apparatus. Note: electrodes E$_{3,4}$ are located above the NbN chip in the $y$ dimension and partially transparent in the figure. (b) Pulsed potential applied to E$_5$ for Rydberg-state-selective electric field ionization. (c) Electron signal recorded at the MCP detector for atoms excited to the $|55\mathrm{s}\rangle$ and $|56\mathrm{s}\rangle$ states, following ionization in the field generated by the potential in (b).\label{fig:setup}}
\end{figure}

The experiments were performed with pulsed supersonic beams of metastable He (mean longitudinal speed $\overline{v}_{\text{He}} = 2000$~m/s; 50~Hz repetition rate)~\cite{morgan20a}. After passing through a 2-mm-diameter skimmer and a charged particle filter, the atomic beam entered the cryogenically cooled central region of the apparatus depicted schematically in Fig.~\ref{fig:setup}(a). Between electrodes E$_1$ and E$_2$ the atoms were excited to the 1s55s\,$^3$S$_1$ level ($|55\mathrm{s}\rangle$) using a the resonance enhanced 1s2s\,$^3$S$_1\rightarrow$1s3p\,$^3$P$_2\rightarrow$1s55s\,$^3$S$_1$ two-photon laser excitation scheme~\cite{hogan18a}. The laser beams were focussed to $\sim50~\mu$m FWHM beam waists to limit the spatial spread of the excited atoms in the $y$ dimension. Rydberg-state photoexcitation occurred for 1.6~$\mu$s, resulting in the preparation of 3.2-mm-long ensembles of excited atoms. 

The Rydberg atoms then traversed a 10~$\times$ 10~mm niobium nitride (NbN) superconducting chip (silicon substrate; 100-nm-thick NbN film; critical temperature $T_{\mathrm{c}}=12.1$~K), where they interacted with the third harmonic microwave field in a $\lambda/4$ superconducting CPW resonator (length 6.335~mm; center-conductor width 20~$\mu$m; insulating-gap width 10~$\mu$m). This resonator was capacitively coupled at the open end to a $\cup$-shaped CPW [see Fig.~\ref{fig:setup}(a)]. The resonator resonance frequency, $\omega_{\mathrm{res}}$, was selected to lie close to the field-free $|55\mathrm{s}\rangle\rightarrow|56\mathrm{s}\rangle$ two-photon transition at $\omega_{55\mathrm{s},56\mathrm{s}}/2=2\pi\times19.556\,499$~GHz (the quantum defects of the $|55\mathrm{s}\rangle$ and $|56\mathrm{s}\rangle$ states are $0.296\,669\,3$ and  $0.296\,668\,8$, respectively~\cite{drake99a}). When the atoms passed above the straight section of the resonator aligned with the atomic beam axis, pulsed potentials were applied to E$_3$ and E$_4$ which were oriented parallel to, and located 10~mm above, the NbN chip surface in the $y$ dimension. This allowed the compensation of stray electric fields at the position of the atoms above the surface. 

After passing the NbN chip, the atoms entered between E$_5$ and E$_6$ where Rydberg-state-selective detection was implemented by applying the pulsed potential in Fig.~\ref{fig:setup}(b) to E$_6$ while E$_5$ was maintained at 0~V. This resulted in a time-varying electric field that ionized the $|56\mathrm{s}\rangle$ state at an earlier time than the $|55\mathrm{s}\rangle$ state. The resulting electrons were accelerated out from the cryogenic part of the apparatus to a microchannel plate (MCP) detector operated at 295~K. The electron signal recorded at the MCP, for atoms prepared in the $|55\mathrm{s}\rangle$ and $|56\mathrm{s}\rangle$ states is displayed in Fig.~\ref{fig:setup}(c). By choosing appropriate detection time windows, the signal from each state could be isolated and monitored.

\begin{figure}
    \includegraphics[width=0.37\textwidth]{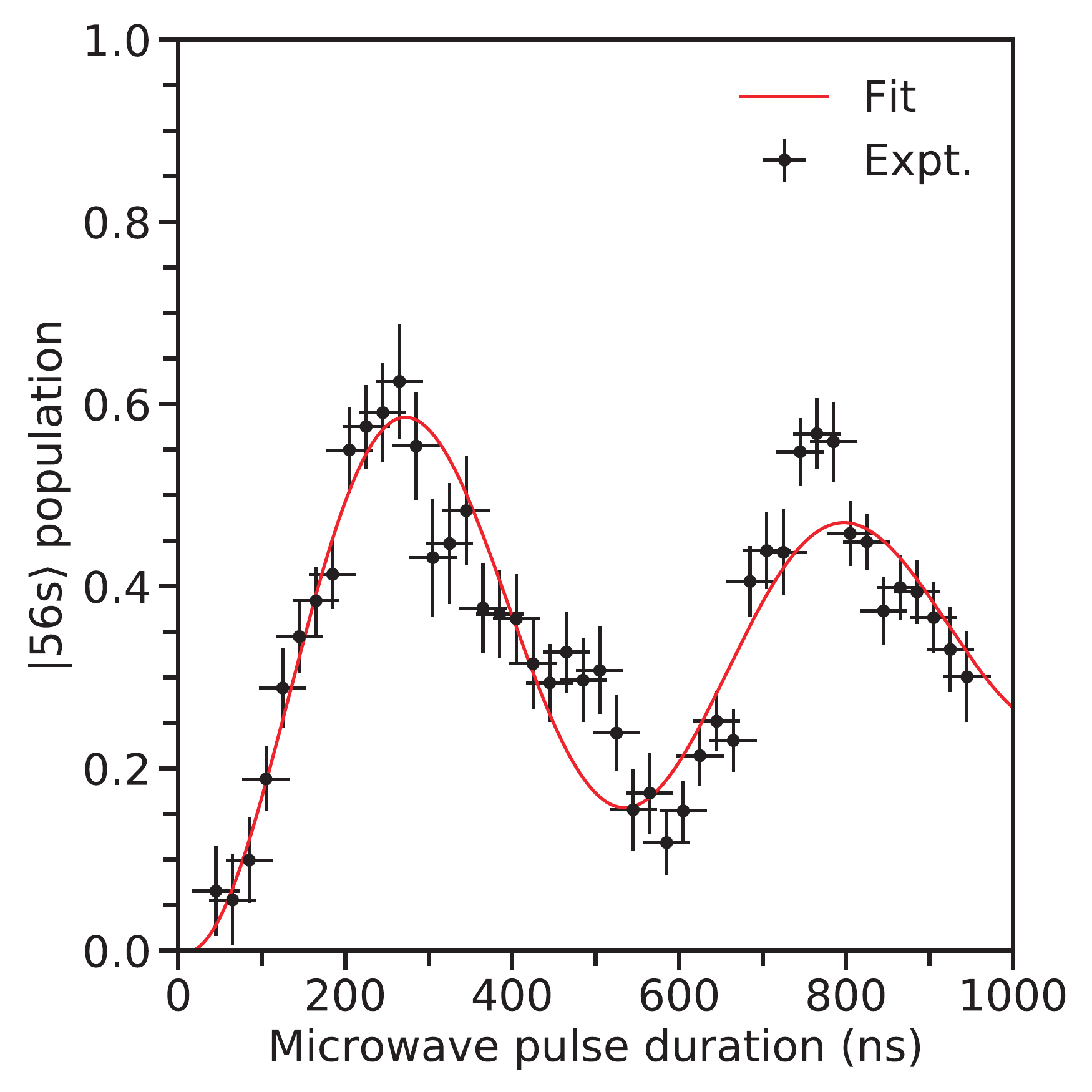}
    \caption{Experimentally recorded (black points) and calculated (continuous red curve) Rabi oscillations in the population of the $|56\mathrm{s}\rangle$ state for $\omega_{\mu}=2\pi\times19.556\,49$~GHz.}
    \label{fig:rabi}
\end{figure}

The coherent interaction of the Rydberg He atoms with the microwave field in the CPW resonator was studied by stabilizing the temperature of the NbN chip to $T_{\mathrm{chip}}=3.9$~K so that the frequency and quality factor of the third harmonic mode, as determined from microwave transmission through the $\cup$-shaped CPW using the circle fit method~\cite{probst15a}, were $\omega_{\mathrm{res}}=2\pi\times(19.556\,12\pm0.02)$~GHz and $Q=2390\pm20$, respectively. The coherence time, $T_2$, of the atom-circuit interface was then determined by injecting a $\omega_{\mu}=2\pi\times19.556\,49$~GHz pulse of microwave radiation into the resonator to drive the two-photon $|55\mathrm{s}\rangle\rightarrow|56\mathrm{s}\rangle$ transition. The duration of this pulse was adjusted while the population of the $|56\mathrm{s}\rangle$ state was monitored, with the results presented in Fig.~\ref{fig:rabi}. Data were recorded with $\sim1$ Rydberg atom prepared in each cycle of the experiment to minimize dephasing arising from fluctuations in collective contributions to the atom-resonator coupling. The value (uncertainty) associated with each data point represents the mean (standard deviation) of 3 repeated measurements each comprising 250 experimental cycles.

The contrast and coherence time of the Rabi oscillations in Fig.~\ref{fig:rabi} result from a detuning $\Delta$ of $\omega_{\mu}$ from $\omega_{55\mathrm{s},56\mathrm{s}}/2$. Estimates of $\Delta$, and the coherence time $T_2$ were obtained from an error-weighted least squares fit of a model function, representing the time-evolution of the population in a two-level atom driven by a microwave field of constant strength and frequency, to the experimental data (continuous red curve in Fig.~\ref{fig:rabi}). Under these conditions the population, $P_{56\mathrm{s}}$, of the $|56\mathrm{s}\rangle$ state is
\begin{eqnarray}
P_{56\mathrm{s}}(t) &=& \frac{\Omega_0^2}{2\Omega^2}\left[1-e^{-t/T_2}\cos\left(\Omega t\right) \right],
\end{eqnarray}
where $\Omega_0$ is the resonant Rabi frequency, and 
$\Omega=\sqrt{\Omega_0^2 + (2\Delta)^2}$. The best-fit function, displayed in Fig.~\ref{fig:rabi}, yielded values of $\Omega_0=2\pi\times(1.57\pm0.04)$~MHz, $\Delta=-2\pi\times(0.54\pm0.01)$~MHz, and $T_2=0.84\pm0.12~\mu$s. This coherence time was limited predominantly by residual shot-to-shot fluctuations in the number of atoms coupled to the resonator field, and their spatial distribution and motion. 

\begin{figure}
    \includegraphics[width=0.37\textwidth]{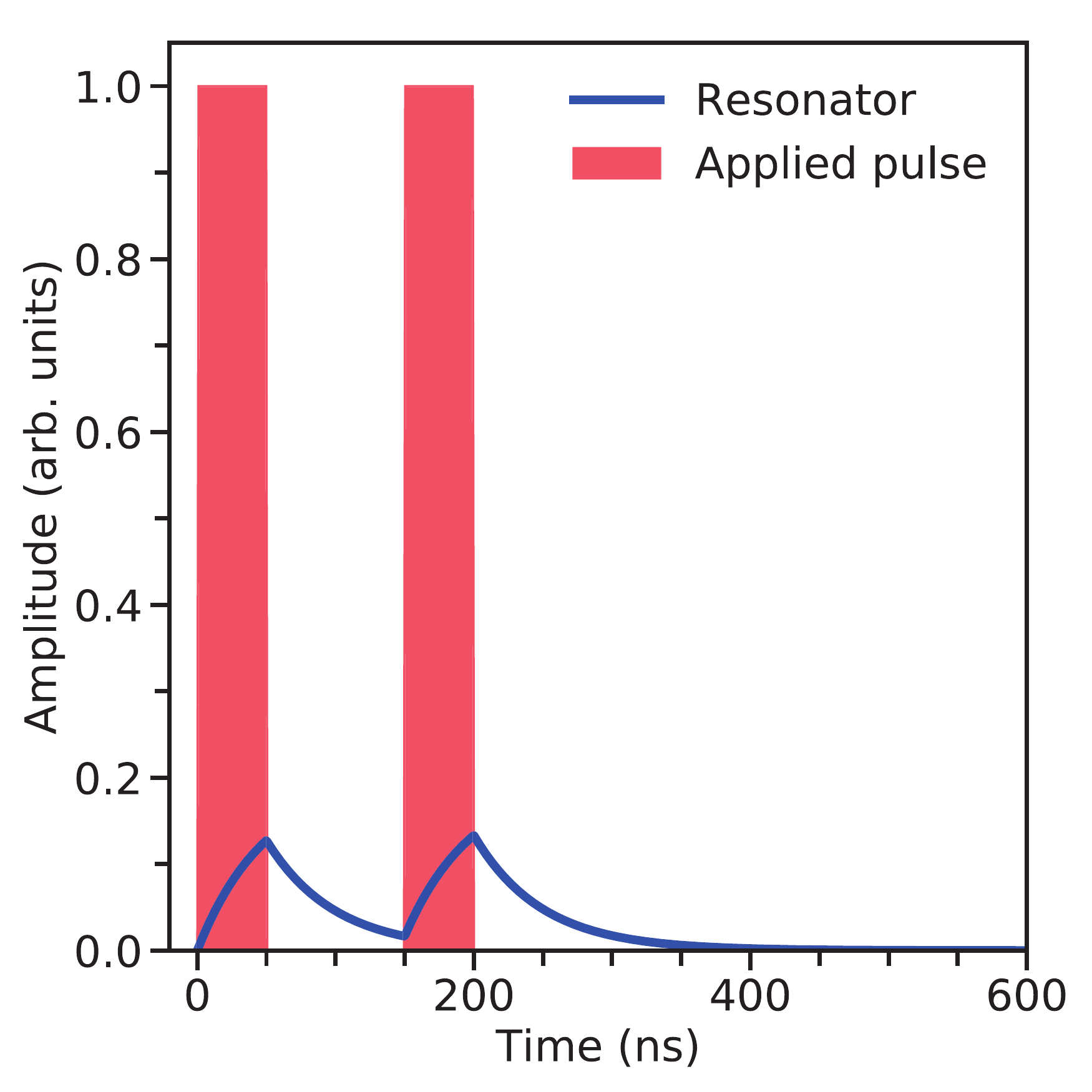}
    \caption{Sequence of microwave pulses (red shaded bands) injected into the CPW resonator for cavity-enhanced Ramsey spectroscopy, and the corresponding calculated time-dependence of the microwave field within the resonator (continuous blue curves).}
    \label{fig:pulses}
\end{figure}

To exploit the atoms in this coherent interface as microscopic quantum sensors to probe the spectral characteristics of the resonator mode, cavity-enhanced Ramsey spectroscopy was performed. In these experiments, a pair of 50-ns-long microwave pulses (Ramsey pulses), separated in time by 100~ns, were injected into the resonator. The FWHM spectral width of these pulses was $\sim2\pi\times17$~MHz, and therefore larger than the $\sim2\pi\times8$~MHz FWHM spectral width of the resonator mode. The time-dependence of the microwave field in the resonator upon injection of these pulses was dominated by the build-up and ring-down in the mode. This can be seen from the calculated response of the resonator field in Fig.~\ref{fig:pulses}, following the propagation of the Ramsey pulses through the CPW. In these calculations the microwave field strength in the resonator, $F_{\mu}(t)$, was evaluated by solving the second-order differential equation 
\begin{equation}
    \frac{\partial^2 F_{\mu}(t)}{\partial t^2} + \frac{\omega_{\mathrm{res}}}{Q}\frac{\partial F_{\mu}(t)}{\partial t} + \omega_{\mathrm{res}}^2 F_{\mu}(t) = \Pi(t)e^{-i\omega_{\mu} t},\label{eq:resonator_response}
\end{equation}
where $\Pi(t)$ is the pulse-sequence envelope. 

\begin{figure}
    \centering
    \includegraphics[width=0.33\textwidth]{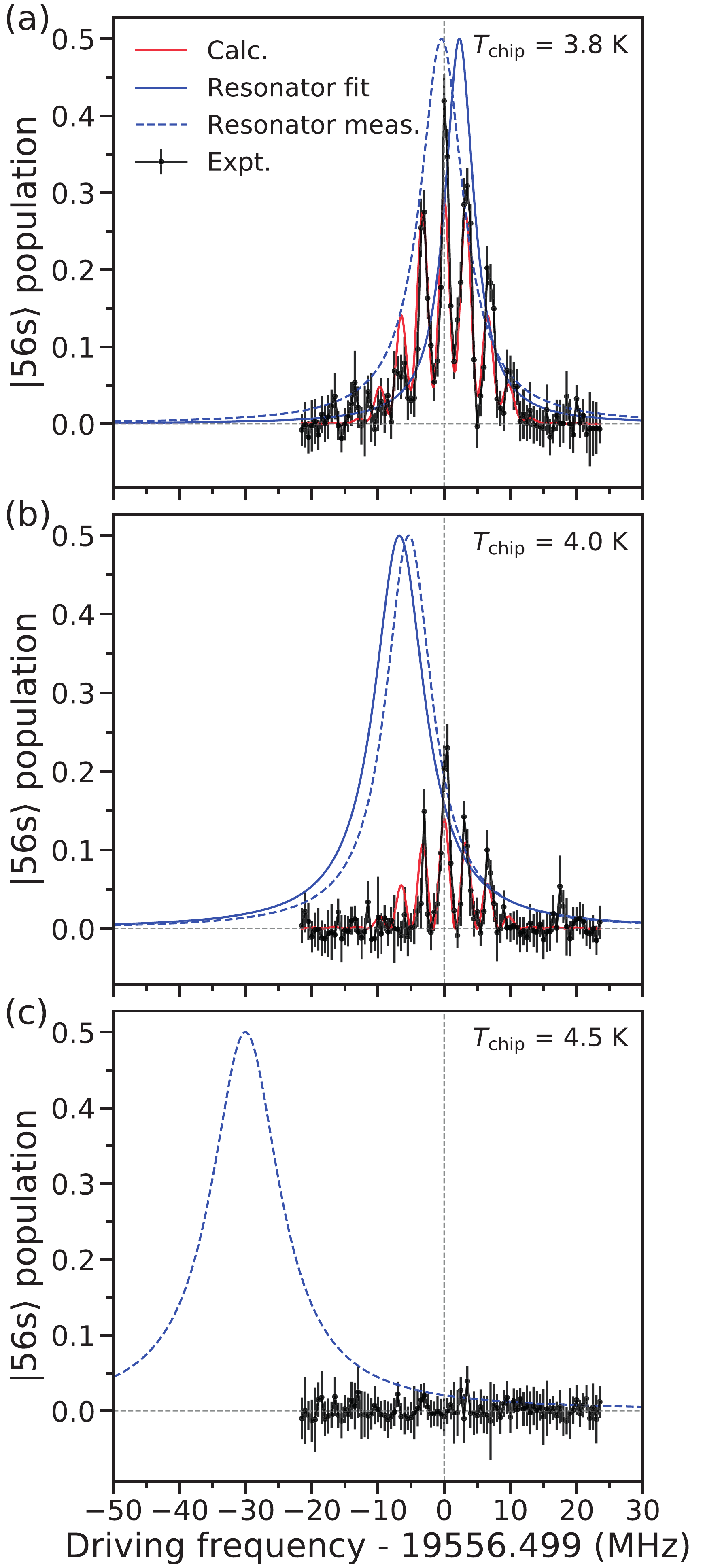}
    \caption{Cavity-enhanced Ramsey spectra of the Rydberg-atom--superconducting-circuit interface. The experimental data (black points) were recorded for (a) $T_{\mathrm{chip}}=3.8$~K, (b) $T_{\mathrm{chip}}=4.0$~K and (c) $T_{\mathrm{chip}}=4.5$~K. The spectral characteristics of the resonator determined from CPW transmission measurements are indicated by the dashed blue curves. The continuous red curves in (a) and (b) represent a spectrum calculated by fitting $\omega_{\mathrm{res}}$ and $Q$ (represented by the continuous blue curves) to the data.}
    \label{fig:ramsey}
\end{figure}

Cavity-enhanced Ramsey spectroscopy was performed with a power of 12~dBm output from the microwave source, which was gated to generate the Ramsey pulses (the attenuation between the source and the NbN chip was -22.5~dB). The contribution from the resonator to the Ramsey spectra was studied with the NbN chip operated at three temperatures, and hence values of $\omega_{\mathrm{res}}$ and $Q$. A first measurement, black points in Fig.~\ref{fig:ramsey}(a), was made for $T_{\mathrm{chip}}=3.8$~K. At this temperature $\omega_{\mathrm{res}}$, as measured by CPW transmission, was detuned from $\omega_{55\mathrm{s},56\mathrm{s}}/2$ by $\sim -2\pi\times0.38$~MHz. The corresponding Lorentzian resonance profile of the resonator mode is indicated by the dashed blue curve. From these data it is seen that the spectral overlap of the resonator mode with the $|55\mathrm{s}\rangle\rightarrow|56\mathrm{s}\rangle$ transition yields a Ramsey spectrum that reflects the resonant enhancement of the microwave field at the interface. The Ramsey fringes in this spectrum have FWHM spectral widths of $2\pi\times(1.295\pm~0.014)$~MHz. However, the amplitude of the spectral profile is not symmetric about the central fringe because of the coupling to the resonator. The incomplete contrast of the fringes reflects the finite $T_2$ coherence time of the interface. 

Additional spectra were recorded by adjusting the value of $T_{\mathrm{chip}}$ to detune the resonator further from $\omega_{55\mathrm{s},56\mathrm{s}}/2$. For $T_{\mathrm{chip}}=4.0$~K [Fig.~\ref{fig:ramsey}(b)], $\omega_{\mathrm{res}}$, measured by CPW transmission, lay $\sim-2\pi\times5.33$~MHz from $\omega_{55\mathrm{s},56\mathrm{s}}/2$. In the corresponding Ramsey spectrum the intensity is reduced compared to that in Fig.~\ref{fig:ramsey}(a) because of the reduced enhancement of the microwave field under these conditions. For $T_{\mathrm{chip}}=4.5$~K [Fig.~\ref{fig:ramsey}(c)] the resonator resonance frequency was detuned by $-2\pi\times30.02$~MHz from $\omega_{55\mathrm{s},56\mathrm{s}}/2$ and, as seen in Fig.~\ref{fig:ramsey}(c), no population transfer to the $|56\mathrm{s}\rangle$ state was observed.

To infer the spectral characteristics of the resonator from the spectra in Fig.~\ref{fig:ramsey}, semiclassical calculations were performed to treat the time-evolution of the Rydberg-state population in the presence of the resonator field. These involved integrating the Bloch equations for a single two-level Rydberg atom coupled by a two-photon transition to the resonator. The differential equation solved had the form
\begin{equation}
    \frac{\partial \vec{r}}{\partial t} = \vec{\mathcal{A}}(t) \times \vec{r}(t) - \vec{\Gamma} \cdot \vec{r}(t),
\end{equation}
with $\vec{\mathcal{A}}(t) = \left(g^*\,\mathbb{R}\text{e}[F_{\mu}(t)^2], g^*\,\mathbb{I}\text{m}[F_{\mu}(t)^2], \omega_{55\mathrm{s},56\mathrm{s}}\right)^{\mathrm{T}}$, where $g^*$ is the normalized atom--resonator-field coupling strength, and $\vec{\Gamma}$ represents the decay rates of the components of the Bloch vector, $\vec{r}$, in the $\sigma_x$, $\sigma_y$, and $\sigma_z$ dimensions.

In these calculations it was assumed that: (1) the $|55\mathrm{s}\rangle$ and $|56\mathrm{s}\rangle$ states represent a two-level system the time evolution of which can be described by a Bloch vector. (2) The evolution of this Bloch vector reflects that of a single ensemble-average two-level atom, and the measured populations correspond to the $\sigma_z$ component. (3) The dynamics of this ensemble-average Bloch vector are driven by the spatially-averaged microwave field experienced by the atoms. (4) The microwave field in the resonator is described by the equations of motion of a damped classical harmonic oscillator. (5) On the experimental timescale ($0-1$~$\mu$s) decoherence of the atom results only from ensemble dephasing and not because of relaxation. And (6) the dephasing of the atom--circuit interface accounts for contributions from the spatial distribution, and motion of the atoms. 

The time evolution of the Rydberg state population was evaluated to calculate Ramsey spectra by specifying the timing of the microwave pulses injected into the resonator, and the values of $\omega_{\mathrm{res}}$, $Q$, $\omega_{55\mathrm{s},56\mathrm{s}}$, and $g^*$. For each set of these parameters, the population of the $|56\mathrm{s}\rangle$ state at the end of the Ramsey sequence was determined. Spectra were then obtained by repeating the calculations for the range of microwave frequencies of interest. 

\begin{table}
\begin{tabular}{ccccc}
\\\toprule
$T_{\mathrm{chip}}$ & $\omega_{\mathrm{res}}^{\mathrm{fit}}$ & $\omega_{\mathrm{res}}^{\mathrm{meas}}$ & $Q^{\mathrm{fit}}$ & $Q^{\mathrm{meas}}$\\
       (K) & ($2\pi\times$MHz) & ($2\pi\times$MHz) &  & \\
    \hline
    3.8 & 19\,558.8$\pm$0.5 & $19\,556.12\pm0.02$ & 3700$\pm$1800 & 2470$\pm$20\\
    4.0 & 19\,549.7$\pm$3.6 & $19\,551.17\pm0.02$ & 2100$\pm$1700 & 2310$\pm$20\\
    \hline
\end{tabular}
\caption{Spectral characteristics of the $\lambda/4$ CWP resonator measured by CPW transmission (meas), and determined by cavity-enhanced Ramsey spectroscopy (fit) for $T_{\mathrm{chip}}=3.8$ and 4.0~K (see text for details).}\label{tab:res} 
\end{table}

To determine $\omega_{\mathrm{res}}$ and $Q$ from the measured spectra, these were chosen, together with $g^*$, to be free parameters in an error-weighted least squares fit. The resulting best-fit spectra, corresponding to the measurements at $T_{\mathrm{chip}}=3.8$ and 4.0~K, are indicated by the red curves in Fig.~\ref{fig:ramsey}(a) and (b), respectively. The corresponding values of $g^*$ were $\sim2\pi\times0.3~$Hz/(V/cm)$^2$. These calculated spectra are in good quantitative agreement with the experimental data. The spectral profile of the resonator obtained by fitting the experimental data is indicated by the continuous blue curve in each panel. The values of $\omega_{\mathrm{res}}$ and $Q$ measured by CPW transmission, and those obtained by fitting the Ramsey spectra -- with uncertainties obtained from the diagonal elements of the covariance matrix -- are summarized in Table~\ref{tab:res}. From these data it is seen that the values of $\omega_{\mathrm{res}}^{\mathrm{fit}}$ obtained from the Ramsey spectra are similar to those measured by CPW transmission, $\omega_{\mathrm{res}}^{\mathrm{meas}}$. The sensitivity of the Ramsey spectra to the value of $Q$ is not so high for resonators of the kind used here with quality factors of $\sim2500$. The quality factors, $Q^{\mathrm{fit}}$, obtained from the Ramsey spectra are similar to those determined by CPW transmission, $Q^{\mathrm{meas}}$, but the uncertainties are larger. 

The results presented in Fig.~\ref{fig:ramsey} demonstrate that cavity-enhanced Ramsey spectroscopy performed at a hybrid Rydberg-atom--superconducting-circuit interface allows the Rydberg atoms to be exploited as microscopic probes of the instantaneous spectral characteristics of the resonator. The characteristics of superconducting CPW resonators depend on the properties of the microwave field injected into them. Since the peak intensity and spectral distribution of the microwave field was different for the CPW transmission measurements performed in a continuous, i.e., cw, mode, and the cavity-enhanced Ramsey spectroscopy performed in a pulsed mode, the differences in the parameters in Table~\ref{tab:res} are not unexpected. In future experiments it will be of interest to employ higher-$Q$ resonators that will cause more pronounced distortions to the spectra. Higher $Q$ factors could be achieved by (1) working at lower microwave frequencies by identifying suitable alternative atomic transitions, (2) operating the resonator at temperatures further below $T_{\mathrm{c}}$, or (3) refining the NbN chip fabrication process. Together, these advances will pave the way for the use of coherent Rydberg-atom--superconducting-circuit interfaces for applications in optical-to-microwave photon conversion, and the implementation of long-coherence-time microwave quantum memories. 

\begin{acknowledgments}
This work was supported by the Engineering and Physical Sciences Research Council under Grant No. EP/L019620/1 and through the EPSRC Centre for Doctoral Training in Delivering Quantum Technologies, and the European Research Council (ERC) under the European Union's Horizon 2020 research and innovation program (Grant No. 683341). The data that support the findings of this study are available from the corresponding author upon reasonable request.
\end{acknowledgments}

\bibliography{cavity_enhanced_ramsey.bib}

\end{document}